\documentclass[aps,prd,onecolumn,groupedaddress,showpacs,nofootinbib,amssymb
]{revtex4}
\usepackage[dvips]{graphicx}
\usepackage{amssymb}
\usepackage{amsmath}
\usepackage{graphicx}
\usepackage{amsfonts}
\usepackage{bm}

\begin{document}

\title{Inflation with a Smooth Constant-Roll to Constant-Roll Era Transition}
\author{
S.D. Odintsov,$^{1,2}$\,\thanks{odintsov@ieec.uab.es} V.K.
Oikonomou,$^{3,4}$\,\thanks{v.k.oikonomou1979@gmail.com}}
\affiliation{ $^{1)}$ICREA, Passeig Luis Companys, 23, 08010 Barcelona, Spain\\
$^{2)}$  Institute of Space Sciences (IEEC-CSIC) C. Can Magrans s/n, 08193 Barcelona, Spain\\
$^{3)}$ Laboratory for Theoretical Cosmology, Tomsk State University
of Control Systems
and Radioelectronics (TUSUR), 634050 Tomsk, Russia\\
$^{4)}$ Tomsk State Pedagogical University, 634061 Tomsk, Russia\\
}

\begin{abstract}
In this paper we study canonical scalar field models with a varying
second slow-roll parameter, that allow transitions between
constant-roll eras. In the models with two constant-roll eras it is
possible to avoid fine-tunings in the initial conditions of the
scalar field. We mainly focus on the stability of the resulting
solutions and we also investigate if these solutions are attractors
of the cosmological system. We shall calculate the resulting scalar
potential and by using a numerical approach, we examine the
stability and attractor properties of the solutions. As we show, the
first constant-roll era is dynamically unstable towards linear
perturbations and the cosmological system is driven by the attractor
solution to the final constant-roll era. As we demonstrate, it is
possible to have a nearly-scale invariant power spectrum of
primordial curvature perturbations in some cases, however this is
strongly model dependent and depends on the rate of the final
constant-roll era. Finally, we present in brief the essential
features of a model that allows oscillations between constant-roll
eras.
\end{abstract}

\pacs{04.50.Kd, 95.36.+x, 98.80.-k, 98.80.Cq,11.25.-w}

\maketitle



\def\pp{{\, \mid \hskip -1.5mm =}}
\def\cL{\mathcal{L}}
\def\be{\begin{equation}}
\def\ee{\end{equation}}
\def\bea{\begin{eqnarray}}
\def\eea{\end{eqnarray}}
\def\tr{\mathrm{tr}\, }
\def\nn{\nonumber \\}
\def\e{\mathrm{e}}

\section{Introduction}

The single canonical scalar field approach for the description of
the inflationary era, has very appealing properties since in some
cases the resulting power spectrum and the scalar-to-tensor ration
are in concordance with the latest Planck data \cite{planck}. For
reviews on this vast research topic, see for example
\cite{inflation1,inflation2,inflation4,inflation5}. However, the
single scalar field models do not predict a sufficient amount of
non-Gaussianities \cite{Chen:2010xka} in the spectrum of primordial
curvature perturbations, and this feature can possibly make the
single scalar models insufficient, if observations reveal
non-Gaussianities. In fact, combined observations of the Cosmic
Microwave Background anisotropy and of the galaxy distribution, may
reveal that the primordial perturbation modes are correlated, so in
effect this would imply that the assumption of a Gaussian
distribution for the primordial modes is wrong. One way to make the
single scalar field models to produce non-Gaussianities, is to
modify the slow-roll condition, and for a recent research stream on
this see
\cite{Inoue:2001zt,Tsamis:2003px,Kinney:2005vj,Tzirakis:2007bf,Namjoo:2012aa,Martin:2012pe,Motohashi:2014ppa,Cai:2016ngx,Motohashi:2017aob,Hirano:2016gmv,Anguelova:2015dgt,Cook:2015hma,Kumar:2015mfa,Odintsov:2017yud},
see also \cite{Lin:2015fqa,Gao:2017uja} for an alternative
viewpoint. The models of Ref.
\cite{Inoue:2001zt,Tsamis:2003px,Kinney:2005vj,Tzirakis:2007bf,Namjoo:2012aa,Martin:2012pe,Motohashi:2014ppa,Cai:2016ngx,Motohashi:2017aob,Hirano:2016gmv,Anguelova:2015dgt,Cook:2015hma,Kumar:2015mfa,Odintsov:2017yud}
are known as constant-roll models \cite{Motohashi:2014ppa}, or by
using the terminology of Ref. \cite{Martin:2012pe}, fast-roll
models, and in the context of these, single scalar field models of
inflation can yield a non-zero amount of non-Gaussianities
\cite{Namjoo:2012aa,Martin:2012pe}.

In a recent work \cite{Odintsov:2017yud}, we demonstrated that by
allowing the second slow-roll index to  smoothly vary as a function
of the scalar field, it is possible to achieve a transition from a
slow-roll era to a constant-roll era. In effect, this would have a
direct impact on the scalar field theory, since a non-zero amount of
non-Gaussianities could be produced during the constant-roll era. In
this paper we shall present how to obtain an inflationary attractor,
which allows the smooth transition between two constant-roll eras.
As we shall demonstrate, for large field values one of the
constant-roll eras is realized, and as the field value decreases,
the smooth transition to the other constant-roll era is achieved. We
shall study in some detail some phenomenological models, and we
shall investigate the stability of the resulting solutions. We
mainly focus on the stability of the inflationary attractor, after
we first validate that the resulting solutions are producing
inflation. Interestingly enough, one of the models we shall present
has a similar potential to the one found in Ref.
\cite{Motohashi:2017aob}. In the analysis that follows, we will show
that the first constant-roll era is dynamically unstable and
therefore it is possible that it is eventually very brief. After the
first constant-roll era, the cosmological system is driven by the
attractor of the system, to the last constant-roll era, which is
dynamically stable towards linear perturbations. With regards to the
dynamical stability, we quantify this study by using a parameter
which measures the deviation from the constant-rolling condition,
and as we show, the perturbations corresponding to the last
constant-roll era are very small, a feature which indicates that the
last era is stable. We critically discuss various implications of
this stability on the models we study and also we briefly address
the primordial curvature perturbations issue. As we show, depending
on the rate of the constant-roll, the power spectrum of the
primordial curvature perturbations may become nearly
scale-invariant, a feature which strongly depends on the rate of
constant-roll.

The motivation for having two constant-roll eras instead of one is
mainly the fact that in the latter case there is less freedom in
choosing the initial conditions for inflation. Particularly, as was
shown in Ref. \cite{Martin:2012pe}, it is necessary to fine-tune the
initial values of the scalar field in order to obtain the desirable
number of $e$-foldings. In the two constant-roll models, this
characteristic is not a necessity anymore, so there is more freedom
to provide various phenomenological aspects.

This paper is organized as follows: In section II, we present the
essential features of the formalism we shall use in this paper, in
section III, we study a class of exponential models that lead to
transition between different constant-roll eras. In section IV, we
discuss a variant form of the model presented in section III, which
allows for the last constant-roll era to be nearly scale invariant.
In section IV we present a model which leads to oscillating
transitions between different de Sitter vacua, and also the
transition between various constant-roll eras is achieved. Finally,
the conclusions follow at the end of the paper.

\section{Essential Features of Dynamical Transition Between Different Constant-Roll Eras}

We assume that the metric describing the spacetime is a flat
Friedmann-Robertson-Walker metric, and also that a canonical scalar
field determines the dynamics, with the action being,
 \be
\label{canonicalscalarfieldaction}
\mathcal{S}=\sqrt{-g}\left(\frac{R}{2}-\frac{1}{2}\partial_{\mu}\phi
\partial^{\mu}\phi -V(\varphi))\right)\, ,
\ee where $V(\varphi)$ is the scalar potential. The corresponding
energy density is,
 \be
\label{energydensitysinglescalar}
\rho=\frac{1}{2}\dot{\varphi}^2+V(\varphi)\, , \ee and in addition
the corresponding pressure is,
 \be \label{pressuresinglescalar}
P=\frac{1}{2}\dot{\varphi}^2-V(\varphi)\, . \ee For the flat
Friedmann-Robertson-Walker metric, the Friedmann equation for the
scalar field is,
 \be
\label{friedmaneqnsinglescalar} H^2=\frac{1}{3 M_p^2}\rho\, , \ee
and we also have,
 \be
\label{dothsinglescalar} \dot{H}=-\frac{1}{2M_p^2}\dot{\varphi}^2\,
. \ee Moreover, the canonical scalar field obeys the Klein-Gordon
equation,
 \be \label{kleingordonsingle}
\ddot{\varphi}+3H\dot{\varphi}+V'=0\, , \ee where the prime denotes
differentiation with respect to  $\varphi$.

The slow-roll parameters $\epsilon$ and $\eta$ control the
inflationary dynamics, and these are the lowest order terms in the
so-called Hubble slow-roll expansion \cite{Liddle:1994dx}, and are
equal to,
 \be \label{slowrollindiceshubblerate}
\epsilon=-\frac{\dot{H}}{H^2}\, ,\quad
\eta=-\frac{\ddot{H}}{2H\dot{H}}\, . \ee In addition, these can be
expressed in terms of the scalar field,
 \be
\label{slowrollindiceshubblerate123}
\epsilon=\frac{\dot{\varphi}^2}{2M_p^2H^2}\, ,\quad
\eta=-\frac{\ddot{\varphi}}{2H\dot{\varphi}}\, . \ee The basic
assumption of the constant-roll models
\cite{Martin:2012pe,Motohashi:2014ppa,Motohashi:2017aob}, is that
the second slow-roll index $\eta$ is not small during the
inflationary era, but it is a constant, that is, $\eta=-n$, with $n$
being a constant. In a recent work we assumed
\cite{Odintsov:2017yud} that the second slow-roll index is of the
form,
\begin{equation}\label{basciccnd1}
\eta=-f(\varphi (t))\, ,
\end{equation}
or equivalently,
\begin{equation}\label{basiccondition}
\frac{\ddot{\varphi}}{2H\dot{\varphi}}=f(\varphi (t))\, .
\end{equation}
The function  $f(\varphi (t))$ appearing above, is assumed to be a
monotonic and smooth function of the scalar field. Due to the fact
that $\dot{H}=\dot{\varphi}\frac{\mathrm{d}H}{\mathrm{d}\varphi}$,
Eq. (\ref{dothsinglescalar}) can be rewritten in the following way,
\begin{equation}\label{extra1}
\dot{\varphi}=-2M_p^2\frac{\mathrm{d}H}{\mathrm{d}\varphi}\, ,
\end{equation}
and by differentiating Eq. (\ref{extra1}) with respect to the cosmic
time $t$, and also by substituting the result in Eq.
(\ref{basiccondition}), we get,
\begin{equation}\label{masterequation}
\frac{\mathrm{d}^2H}{\mathrm{d}\varphi^2}=-\frac{1}{2M_p^2}f(\varphi
)H(\varphi )\, .
\end{equation}
The above differential equation will determine the Hubble rate as a
function of the canonical scalar and the resulting solution should
be checked explicitly if it is the attractor of the cosmological
system. This general strategy of finding the solution $H(\varphi)$
was developed in Ref. \cite{Lidsey:1991zp,Lidsey:1991dz}, and we
adopt this strategy in the present article. The scalar potential in
terms of $H(\varphi)$ is equal to,
\begin{equation}\label{potentialhubblscalar}
V(\varphi)=3M_p^2H(\varphi)^2-2M_p^4(H'(\varphi))^2\, .
\end{equation}
As we already mentioned, a solution $H_0(\varphi)$ of the
differential equation (\ref{masterequation}) is not necessarily an
attractor of the cosmological equations, so this must be checked
both numerically and if possible analytically. With regards to the
analytic approach, we can follow the following procedure: Consider
the variation of Eq. (\ref{potentialhubblscalar}), which is,
\begin{equation}\label{perturbationeqnsbbasic1}
H_0'(\varphi)\delta H'(\varphi)\simeq
\frac{3}{2M_p^2}H_0(\varphi)\delta H(\varphi)\, ,
\end{equation}
so for a given solution $H_0(\varphi)$, we get,
\begin{equation}\label{perturbationsolution1}
\delta H(\varphi )=\delta
H(\varphi_0)e^{\frac{3}{2M_p^2}\int_{\varphi_0}^{\varphi}\frac{H_0(\varphi)}{H_0'(\varphi
)}\mathrm{d}\varphi} \, ,
\end{equation}
where $\varphi_0$ some initial value of the  canonical scalar field.
Hence, if the linear perturbations (\ref{perturbationeqnsbbasic1})
decay for the solution $H_0(\varphi)$, then, the solution
$H_0(\varphi)$ is stable, and in the opposite case, the solution is
unstable. The stability of the solution towards linear
perturbations, indicates that the solution is an attractor of the
theory, since all the solutions converge to $H_0(\varphi).$
\begin{figure}[h]
\centering
\includegraphics[width=15pc]{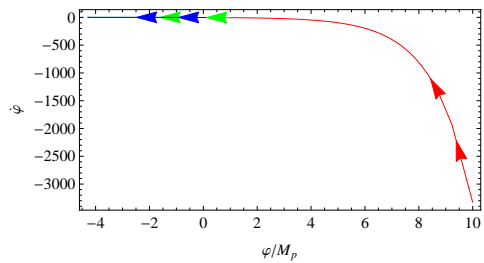}
\caption{Phase space structure of the solution
(\ref{generalsolution1}), with the following initial conditions:
$\varphi (0)=10M_p$ (red line), $\varphi (0)=M_p$ (green line),
$\varphi (0)=0.1M_p$ (blue line).}\label{plotsnumerics1}
\end{figure}
In addition to this analytic approach, the structure of the phase
space $(\dot{\varphi}(t),\varphi (t))$, which can be found
numerically in most cases, will reveal if a solution of the
differential equation (\ref{masterequation}) is an attractor of the
cosmological equations. In the following two sections we shall
present to models that realize a constant-roll to constant-roll
transition, and we shall investigate if the resulting solutions
$H(\varphi )$ are stable and if these are the attractors of the
cosmological equations.
\begin{figure}[h]
\centering
\includegraphics[width=15pc]{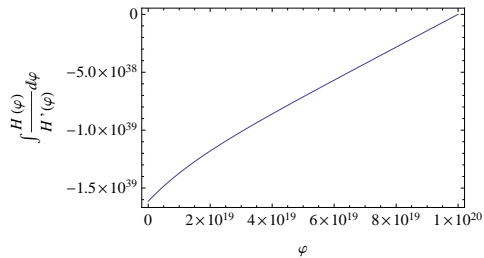}
\caption{Stability of linear perturbations for the solution
(\ref{generalsolution1}).}\label{stabilityperturbations}
\end{figure}

\section{Model I}

Let us assume that the function $f(\varphi (t))$ in Eq.
(\ref{basiccondition}) has the following form,
\begin{equation}\label{choice1}
f(\varphi )=-\frac{\beta  \exp (\lambda  \varphi)}{\delta +\beta  \exp
(\lambda  \varphi)}\, ,
\end{equation}
which is a monotonic function with respect to $\varphi $. The
parameters $\beta$ and $\delta$ are dimensionless positive
constants, and the parameter $\lambda$ will be chosen as follows,
\begin{equation}\label{yperif}
\lambda =\frac{1}{\sqrt{2}M_p}\, .
\end{equation}
For $\frac{\varphi}{M_p}\gg 1$, then the function $f(\varphi)$
behaves as $f(\varphi)\sim -1$, so this describes a constant-roll
era. Particularly, by using the notation of
\cite{Motohashi:2014ppa}, this corresponds to $\alpha=-2$.

As the field values drop, and when $\frac{\varphi}{M_p}\ll 1$ the
function $f(\varphi)$ behaves as $f(\varphi) \simeq -\frac{\beta
}{\beta +\delta }$. Therefore, by appropriately choosing the
parameters, various scenarios with constant-roll can be described.
Here we focus on the case with $\beta=\delta$, but also alternative
scenarios can be described by choosing $\beta$ and $\delta$
differently. Before we discuss the transition from one constant-roll
era to the other, we will focus on the stability of the solution
$H(\varphi)$ corresponding to the model (\ref{choice1}).
Particularly, we are interested in showing that an the resulting
solution $H(\varphi)$ is an attractor of the cosmological system. By
solving the differential equation (\ref{masterequation}), we get the
following solution,
\begin{equation}\label{generalsolution1}
H(\varphi)=C_1\, \delta +\beta  \exp (\lambda  \varphi)\, ,
\end{equation}
where $C_1$ is an integration constant. Note that the full solution
of the differential equation (\ref{masterequation}) for the choice
(\ref{choice1}) is the following,
\begin{equation}\label{asxref1}
H(\varphi)=C_1\, \delta +\beta  \exp (\lambda  \varphi)+C_2
\left(\frac{\beta  e^{\lambda  \varphi } \log \left(\frac{\delta
e^{-\lambda  \varphi }}{\beta }+1\right)}{\delta }+\log
\left(\frac{\delta  e^{-\lambda  \varphi }}{\beta
}+1\right)-1\right)\, ,
\end{equation}
where $C_2$ is an additional integration constant. So we assume that
$C_2=0$, in order for the constant-roll transition to occur. In
principle the existence of a non-zero $C_2$ will alter the physical
picture we describe in this paper, so we do not discuss this case.
By substituting the resulting $H(\varphi)$ of Eq.
(\ref{generalsolution1}) in Eq. (\ref{potentialhubblscalar}), we
obtain the scalar potential,
\begin{equation}\label{potentialcase1}
V(\varphi)=2 \beta ^2 M_p^2 e^{\frac{\sqrt{2} \varphi }{M_p}}+6
\beta  \delta  M_p^2 e^{\frac{\varphi }{\sqrt{2} M_p}}+3 \delta ^2
M_p^2\, ,
\end{equation}
where we used (\ref{yperif}). We can express the Hubble rate as a
function of the cosmic time as follows: by substituting $H(\varphi)$
from Eq. (\ref{generalsolution1}) in the differential equation
(\ref{dothsinglescalar}), and by solving this we can find the
function $\varphi (t)$. Hence by substituting the resulting $\varphi
(t)$ back in Eq. (\ref{generalsolution1}), we can obtain the
function $H(t)$, which is,
\begin{equation}\label{ht1}
H(t)=\delta +\frac{1}{t}\, .
\end{equation}
Hence, the second derivative of the scale factor $\ddot{a}$,
satisfies $\ddot{a}>0$. Let us investigate whether the solution
(\ref{generalsolution1}) is a stable attractor of the theory, and in
Fig. \ref{plotsnumerics1} we have plotted the phase space behavior
of the solution (\ref{generalsolution1}), by using the initial
conditions $\varphi (0)=10M_p$ (red line), $\varphi (0)=M_p$ (green
line), $\varphi (0)=0.1M_p$ (blue line), for $\beta=\delta=2$. We
need to note that for the numerical analysis we numerically solve
the differential equation of Eq. (\ref{extra1}), which we quote here
for reading convenience,
\begin{equation}\label{extra1additionalref}
\dot{\varphi}(t)=-2M_p^2\frac{\mathrm{d}H(\varphi
(t))}{\mathrm{d}\varphi}\, ,
\end{equation}
which needs to be solved as a function of the cosmic time $t$, and
the function $H(\varphi (t))$ must be chosen as in Eq.
(\ref{generalsolution1}). Since the differential equation is of
first order, only one initial condition is needed, this is why we
specified only the value of the scalar field at various cosmic time
instances. So the numerical analysis will reveal if the solution
(\ref{generalsolution1}) is a stable attractor of the
one-dimensional dynamical system appearing in Eq.
(\ref{extra1additionalref}). Note that the explicit form of the
function $H(t)$ is not needed if we use the dynamical system
(\ref{extra1additionalref}). As it can be seen in Fig.
\ref{plotsnumerics1}, the phase space has an attractor which is
reached by all the solutions as the field values decrease and also
when the velocity of the field decreases too. Hence, the solution
(\ref{generalsolution1}) is an attractor and this can also be
verified analytically. By using Eq. (\ref{perturbationsolution1}),
for the solution (\ref{generalsolution1}), we obtain,
\begin{equation}\label{integrationanalytic}
\int_{\varphi}^{\varphi}\frac{H(\varphi)}{H'(\varphi)}\mathrm{d}\varphi=-\frac{\beta
e^{-\lambda \varphi }-\beta e^{-\lambda  \varphi_i}+\delta  \lambda
({\varphi_i}-\varphi )}{\delta \lambda ^2}\, ,
\end{equation}
which is negative for the field values of interest. This can also be
seen in Fig. (\ref{stabilityperturbations}), where we plotted the
behavior of the perturbation exponent. Actually, the fact that the
quantity appearing in Eq. (\ref{integrationanalytic}) is negative is
an important feature, as it was also pointed in Ref.
\cite{Liddle:1994dx}, but also the important point is that the decay
of the perturbation is exponential, as it can be seen from Eq.
(\ref{perturbationsolution1}). Hence, we have verified that it is
possible to realize the constant-roll to constant-roll transition,
for the model (\ref{choice1}), and the resulting solution
(\ref{generalsolution1}) is a stable attractor of the theory.

Note that the numerical analysis we performed above aimed to
investigate whether the solution $H(\varphi)$, appearing in Eq.
(\ref{generalsolution1}), is the stable attractor of the
cosmological system, therefore the potential of Eq.
(\ref{potentialcase1}) does not affect the phase space diagram, at
least in our approach. As it can be seen in Fig.
\ref{plotsnumerics1}, there appear three curves which converge to
the attractor point in the upper left. The difference between the
three curves is that the red curve starts around $\varphi/M_p\sim
10$, the green one starts at $\varphi/M_p\sim 1$ and also the blue
at $\varphi/M_p\sim 0.1$, as it is expected from the initial
conditions. Recall that the dynamical system that corresponds to the
phase space diagram is that of Eq. (\ref{extra1additionalref})
appearing above, so the solution (\ref{generalsolution1}) is the
attractor of the cosmological dynamical system
(\ref{extra1additionalref}). However, as we show later on, not all
constant-roll models result to this behavior (see Fig. 10 in a later
section).

Now let us investigate how the actual transition of one
constant-roll era to the other is achieved, by distinguishing the
two different eras, that is, we start with the era with $\eta= 1$
and we shall try to see when this era ends, and also how it is
possible for this era to end, so that the era with $\eta=1/2$
starts. It is conceivable that the continuous transition is
guaranteed by the solution $H(\varphi)$ we found in Eq.
(\ref{generalsolution1}), but if we distinguish the limiting cases
of constant roll, then it is qualitatively more easy to understand
some fundamental elements of the evolution, like for example the
growth of perturbations. From a qualitative point of view, a
constant-roll era can end if some sort of instability is caused in
the cosmological system. In our case the transition is ensured by
the form of the potential and of the function $H(\varphi)$, and it
is smooth, but it is worth discussing how this is possible by
distinguishing the two eras.
\begin{figure}[h]
\centering
\includegraphics[width=15pc]{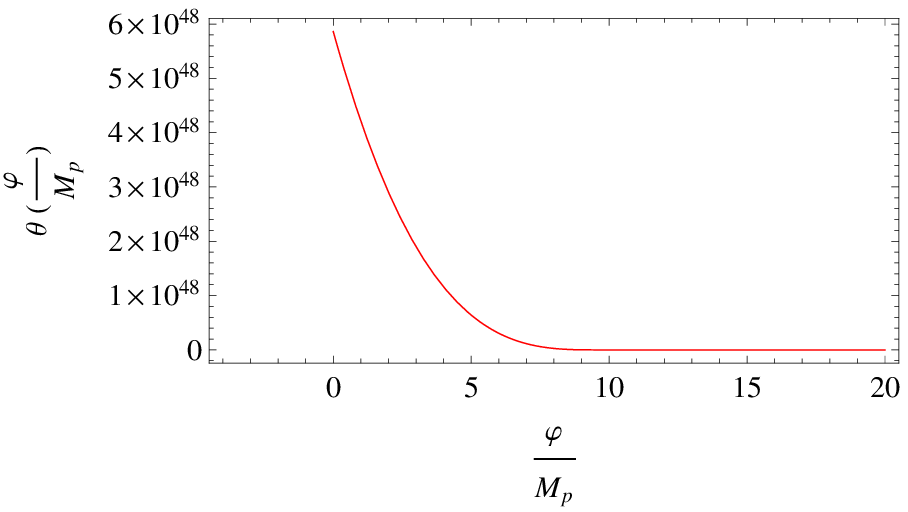}
\includegraphics[width=15pc]{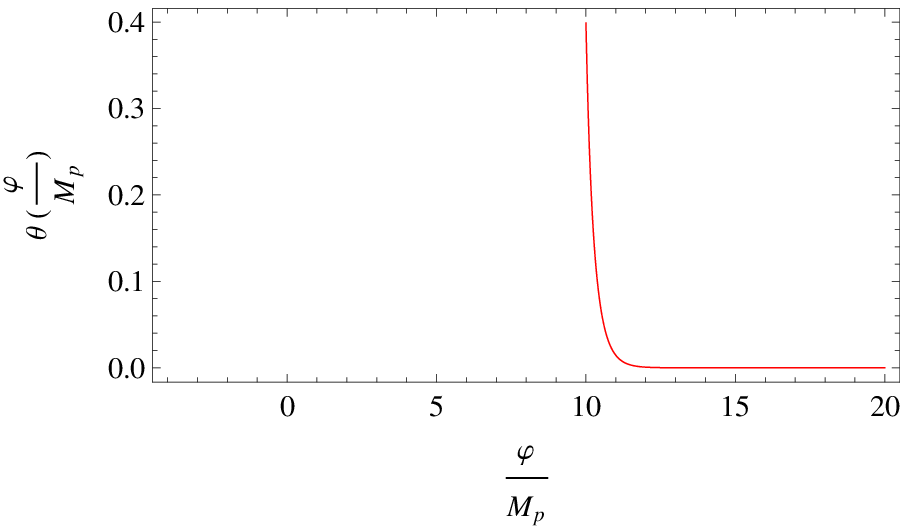}
\includegraphics[width=15pc]{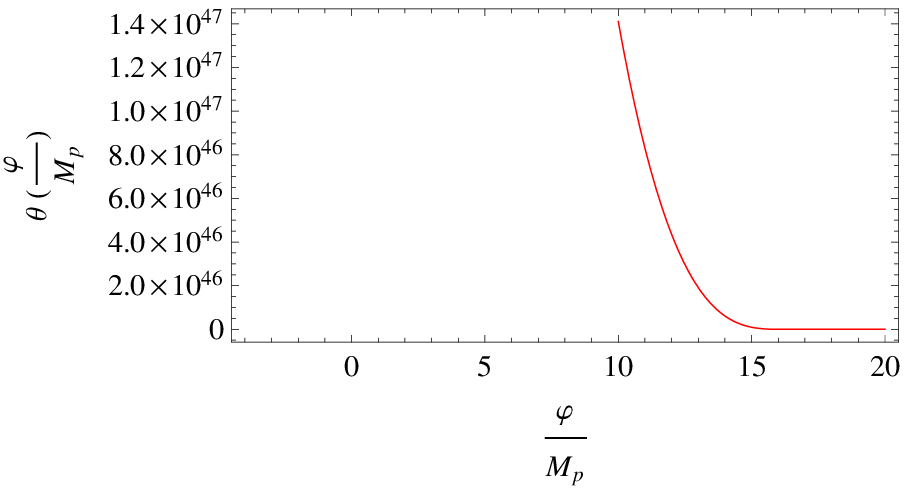}
\caption{Behavior of the perturbations $\theta (\varphi )$ for the
constant-roll era with
$\frac{\ddot{\varphi}}{\dot{\varphi}H}=-1$.}\label{newfigs}
\end{figure}
So let us discuss the general case first, and we will perturb the
limiting case,
\begin{equation}\label{generalcasen}
\frac{\ddot{\varphi}}{\dot{\varphi}H}=n\, .
\end{equation}
Effectively we will examine the behavior of the perturbations around
the limiting case where $f(\varphi)\sim n$, with $n$ being a real
number. As in Ref. \cite{Martin:2012pe}, we define the following
quantity,
\begin{equation}\label{fquantity}
\Theta (\varphi )=\frac{\ddot{\varphi}}{n\dot{\varphi}H}\, ,
\end{equation}
and a study of the evolution of $\Theta(\varphi )$ as a function of
the scalar field $\varphi$ may reveal when does the constant-roll
era (\ref{generalcasen}) ends. Notice that the following holds true,
\begin{equation}\label{holdstrue}
\Theta (\varphi )=\frac{f(\varphi)}{\varphi}\, ,
\end{equation}
which shows that the variable $\Theta (\varphi)$ is strongly
affected by the function $f(\varphi )$. This means that when the
function $f(\varphi)$ strongly varies as a function of $\varphi$,
then the function $\Theta (\varphi)$ also varies strongly. Actually,
if we consider the linear perturbation of the solution $\Theta
(\varphi)$, when the perturbations start to grow, which may occur
after a particular value of the scalar field is reached, then the
constant-roll era (\ref{generalcasen}) practically ends. By
differentiating the function $\Theta (\varphi)$ with respect to
$\varphi$, and by using the following relations,
\begin{align}\label{difftheresultingeqnnrelations}
&
\epsilon_1=\frac{2M_p^2}{H^2}\frac{\mathrm{d}^2H}{\mathrm{d}\varphi^2},\,\,\,\frac{\mathrm{d}\ddot{\varphi}}{\mathrm{d}\varphi}=-V''-3\dot{H}-\frac{3H\ddot{\varphi}}{\dot{\varphi}},\,\,\,
\end{align}
after some algebra, it can be shown that the evolution of the
quantity $\Theta (\varphi)$ is determined by the following equation,
\begin{equation}\label{insertperteqn1}
\frac{\mathrm{d}\Theta
(\varphi)}{\mathrm{d}\varphi}=-\frac{V''(\varphi)}{n H
(\varphi)\dot{\varphi}}+\frac{3
H(\varphi)\epsilon_1}{n\dot{\varphi}}-\frac{3 H(\varphi)\Theta
(\varphi)}{\dot{\varphi}}+\frac{\Theta
(\varphi)H(\varphi)\epsilon_1}{\dot{\varphi}}-\frac{\Theta
(\varphi)^2H(\varphi)n}{\dot{\varphi}}\, ,
\end{equation}
where $\epsilon_1$ is the first slow-roll index. Obviously, during
the constant-roll era (\ref{generalcasen}), the quantity $\Theta
(\varphi)$ is equal to one, so we linearly perturb the solution
$\Theta (\varphi)=1+\theta (\varphi)$, and the perturbation $\theta
(\varphi)$ satisfies the following differential equation,
\begin{equation}\label{equationperturbations2}
\frac{\mathrm{d}\theta
(\varphi)}{\mathrm{d}\varphi}=-\frac{H(\varphi)\theta
(\varphi)\left( -3-3n(2+\theta
(\varphi))+\frac{2M_p^2}{H(\varphi)^2}\frac{\mathrm{d}^2H(\varphi)}{\mathrm{d}\varphi^2}\right)}{2M_p^2\frac{\mathrm{d}H(\varphi)}{\mathrm{d}\varphi}}\,
.
\end{equation}
The function $\theta (\varphi)$ captures the perturbations around
the limiting constant-roll era (\ref{generalcasen}), and in order to
obtain the differential equations (\ref{insertperteqn1}) and
(\ref{equationperturbations2}) we did not assume that the condition
 (\ref{generalcasen}) holds true, this is just the limiting case
 scenario. The impact of the $f(\varphi )$ on the differential
 equations is imprinted in all cases on the variables $\Theta
 (\varphi)$ and $\theta (\varphi)$, and mainly on the latter, since
 this is the perturbation of the limiting case (\ref{generalcasen}).
 In addition, it is the solution (\ref{generalsolution1}) which will
 determine the evolution of the perturbations and recall that the
 solution (\ref{generalsolution1}) was obtained by assuming that the
 function $f(\varphi) $ is not constant, but it is the one given in
 Eq. (\ref{choice1}).

So by having the solution $H(\varphi)$ at hand, given in Eq.
(\ref{generalsolution1}), we can investigate how the perturbations
$\theta (\varphi)$ evolve as a function of the scalar field
$\varphi$. Recall that the constant-roll limit of the function
$f(\varphi)$ for large $\varphi$ is $f(\varphi)=-1$, so in the large
field limit we have $n=-1$. Therefore we shall perform a numerical
analysis of the differential equation
(\ref{equationperturbations2}), with $H(\varphi)$ being as in Eq.
(\ref{generalsolution1}). In Fig. \ref{newfigs} we present the
results of our numerical analysis, for which we used various initial
conditions. The resulting behavior of the perturbations $\theta
(\varphi)$ strongly depend on the choice of the initial condition
for $\theta (\varphi)$, which must be chosen in such a way so that
at large field values, we must have $\theta (\varphi)\ll 1$, in
order to have $\Theta (\varphi)\simeq 1$. In Fig. \ref{newfigs} the
plots are done in terms of $\varphi/M_p$, so the upper left and
right plots correspond to $\theta (20)\simeq 10^{-13.7}$, while the
bottom plot corresponds to $\theta (20)\simeq 10^{-5}$. It is
conceivable that from a physical point of view, the perturbations
should grow, since the function $f(\varphi)$ at $\varphi/M_p$ still
depends on the scalar field and decays to the final value
$f(\varphi)\sim -1/2$ as the field values decrease. So in principle
small perturbations will destabilize the system, as it is expected.
This behavior can be seen in all the plots of Fig. \ref{newfigs}.
The parameter $\Theta (\varphi )=f(\varphi )/n$ changes its values
quite slowly from $\varphi/M_p\sim 20$  until $\varphi/M_p\sim 10$,
so the perturbations, even small, destabilize the initial solution
at $\varphi/M_p\sim 10$, as it can be seen in the upper right plot,
or even at $\varphi/M_p\sim 15$, as it is shown in the bottom plot.
Practically, the $n=-1$ era lasts some $e$-foldings, as it was
expected, and the system makes a transition to intermediate states
in the interval $n=[-1,-1/2]$. Eventually for $\varphi/M_p \preceq
1$, the final constant-roll era with $n=-1/2$ is reached.

\begin{figure}[h]
\centering
\includegraphics[width=15pc]{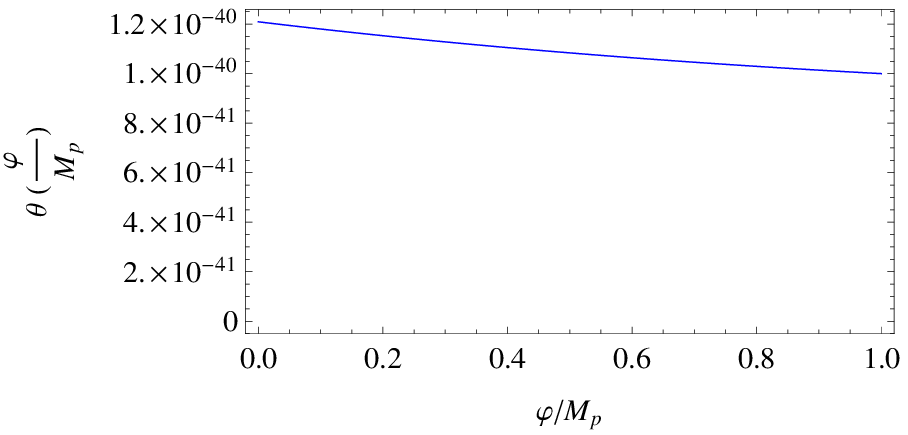}
\includegraphics[width=15pc]{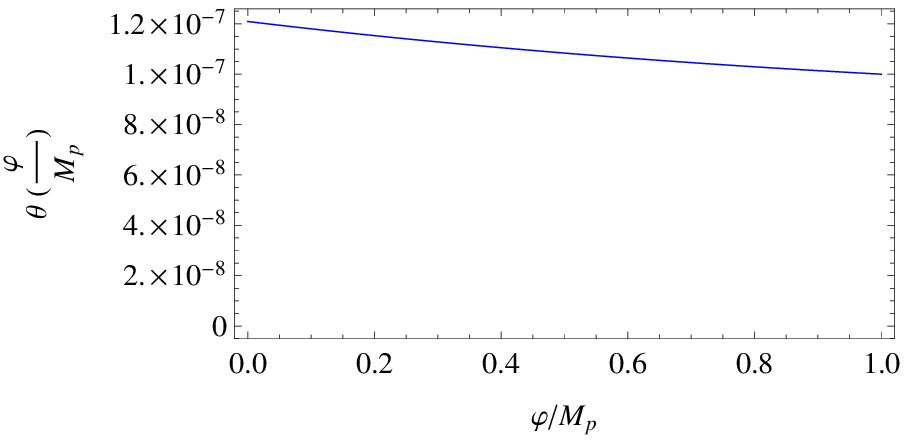}
\caption{Behavior of the perturbations $\theta (\varphi )$ for the
constant-roll era with
$\frac{\ddot{\varphi}}{\dot{\varphi}H}=-1/2$.}\label{newfigs1}
\end{figure}

By following the same procedure, we can investigate what happens
when the second constant-roll era is reached, which corresponds to
$\varphi/M_p\preceq 1$. As it can be seen in Fig. \ref{newfigs1},
the numerical solution for the evolution of the perturbation $\theta
(\varphi/M_p)$ seems quite stable, regardless what initial
conditions we choose. For example, in the left plot of Fig.
\ref{newfigs1} we chose $\theta (1)\sim 10^{-5}$ and for the right
plot $\theta (1)\sim 10^{-40} $, and in both cases the perturbations
remain quite small.
The behavior described by Fig. \ref{newfigs1} was expected, since
the function $\Theta (\varphi)$ does not evolve after the
constant-roll era $n=-1/2$ is reached. This now raises the issue of
graceful exit for this particular case, but this feature strongly
depends on the choice of the parameters $\beta$ and $\delta$, and
also on the choice of the function $f(\varphi)$ which controls the
smooth transition between the constant-roll eras. However, we did
not aim to provide a completely viable model, but just to present
how the transition mechanism behaves and how the actual transition
can be achieved, mainly focusing on the qualitative features of it.
Another question we would like to address in brief is the primordial
curvature perturbations issue, and in the literature there appear
already standards approaches devoted on the constant-roll era, see
for example \cite{Martin:2012pe,Motohashi:2014ppa}. Actually the
results of \cite{Motohashi:2014ppa} are identical to the ones
obtained in \cite{Martin:2012pe}, but the notation is different. In
both cases an important assumption which we will also adopt is that
the power spectrum must not be calculated at the horizon crossing,
but at the time of consideration, possibly when inflation comes to
an end. Also an important assumption is that the subhorizon state is
described by a Bunch-Davies vacuum state, but this is not
necessarily true in the case at hand, however we do not discuss this
issue further at this point.

Coming back to the power spectrum, we quote here the result of
\cite{Martin:2012pe,Motohashi:2014ppa} for the power spectrum, which
is,
\begin{equation}\label{pz}
\mathcal{P}_{\mathcal{R}}(k)\sim k^{-2n}\, ,
\end{equation}
and the tilt of the power spectrum is,
\begin{equation}\label{toilt}
n_s-1\equiv\frac{d\ln\mathcal{P}_{\mathcal{R}}}{d\ln k}\, ,
\end{equation}
so the resulting spectral index is,
\begin{equation}\label{ns1}
n_s=1-2n\, .
\end{equation}
However, for the case $n=-1/2$ the power spectrum is not scale
invariant, as it can be easily seen. Also the resulting spectral
index which is $n_s$ is not compatible with the observational data
coming from the Planck collaboration \cite{planck}. Of course our
aim was not to provide a completely viable model, but to show how
the transition mechanism between constant-roll eras works. If the
function $f(\varphi)$ is chosen to have better behavior than the
particular choice we made, then perhaps a more refined phenomenology
could in principle be produced. Such an example is given in the next
section.

\section{Model II}

Let us discuss the qualitative features of another model, in which
case the function $f(\varphi (t))$ of Eq. (\ref{basiccondition}) has
the following form,
\begin{equation}\label{choice1thrylos}
f(\varphi )=-(\beta ^2+\alpha  e^{ -\lambda  \varphi})\, ,
\end{equation}
which is a monotonic function of $\varphi $. The parameter $\lambda$
is chosen as in Eq. (\ref{yperif}), while $\alpha$ and $\beta$ are
positive dimensionless real numbers, which we specify later on in
this section.

\begin{figure}[h]
\centering
\includegraphics[width=15pc]{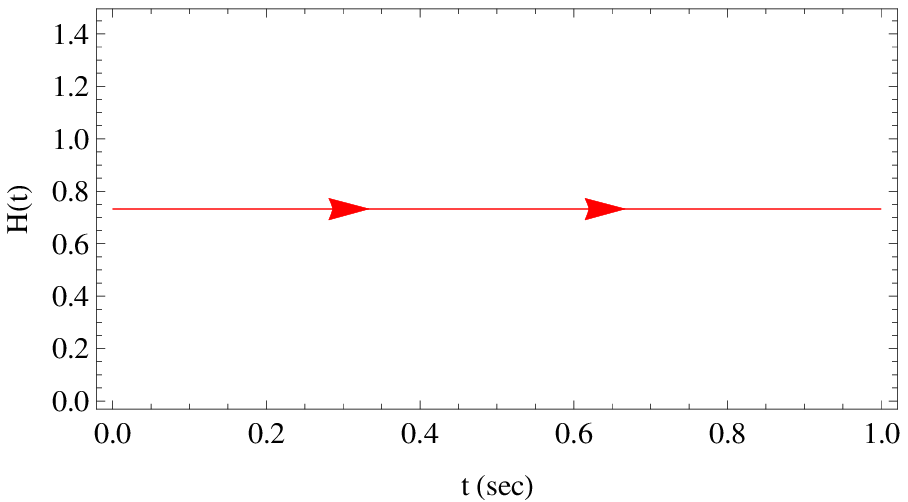}
\includegraphics[width=15pc]{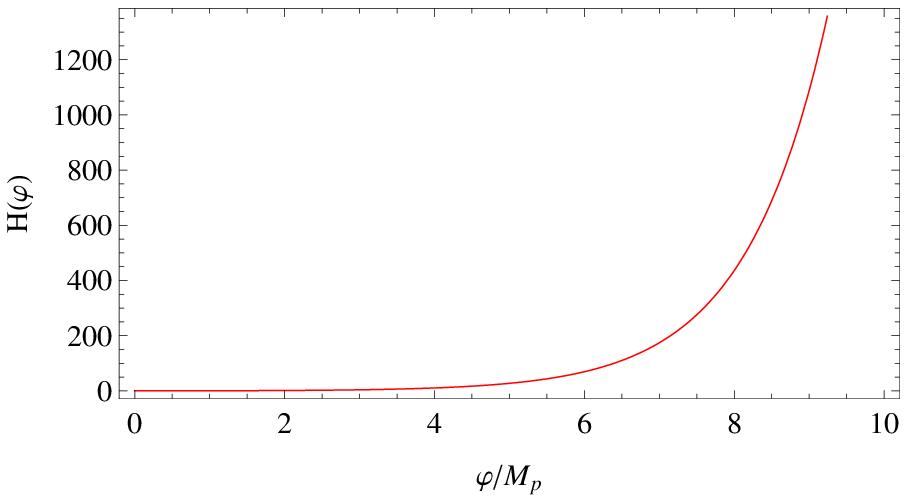}
\caption{The Hubble rate as a function of the cosmic time (left) and
as a function of the scalar field $\varphi/M_p$.}\label{hubplot123}
\end{figure}

\begin{figure}[h]
\centering
\includegraphics[width=15pc]{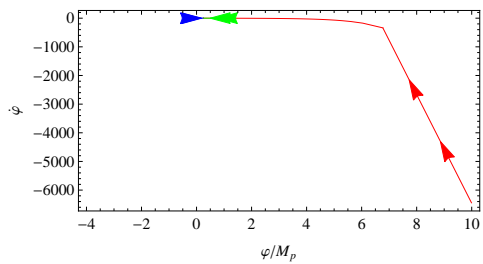}
\caption{The phase space of the solution $H(\varphi)=C_1
I_{-\beta}(2 \sqrt{\alpha } e^{-\frac{1}{2} (\lambda x)})$, for the
initial conditions $\varphi (0)=10M_p$ (red line), $\varphi (0)=M_p$
(green line), $\varphi (0)=0.1M_p$ (blue
line).}\label{hargarthrylos}
\end{figure}

In the limit $\frac{\varphi}{M_p}\gg 1$, the function $f(\varphi)$
behaves as $f(\varphi)\sim -\beta^2$, so this describes a
constant-roll era, and in the limit $\frac{\varphi}{M_p}\ll 1$ the
function $f(\varphi)$ behaves as $f(\varphi) \simeq -(\beta
^2+\alpha )$. A phenomenologically interesting scenario occurs for
the following choice of the parameters,
\begin{equation}\label{choiceofparameterslocal}
\alpha=1.35,\,\,\,\beta=\sqrt{\frac{5}{3}}\, ,
\end{equation}
in which case in the large field limit, we have $f(\varphi)\sim
-5/3$, while in the small field limit we have $f(\varphi)\sim
-3.0167$. By solving the differential equation
(\ref{masterequation}), in this case we get,
\begin{equation}\label{generalsolution1thrylos}
H(\varphi)=C_1 I_{-\beta}(2 \sqrt{\alpha } e^{-\frac{1}{2} (\lambda
x)})\, ,
\end{equation}
where $C_1$ is an integration constant. The corresponding scalar
potential can be found by using (\ref{potentialhubblscalar}),
\begin{align}\label{potentialcase1thrylos}
& V(\varphi)=-\frac{1}{2} \alpha  C_1^2 \lambda ^2 M_p^4 e^{-\lambda
\varphi } I_{-2 \beta -1}\left(2 e^{-\frac{1}{2} (\lambda  \varphi
)} \sqrt{\alpha }\right){}^2-\frac{1}{2} \alpha C_1^2 \lambda ^2
M_p^4 e^{-\lambda  \varphi } I_{1-2 \beta }\left(2
e^{-\frac{1}{2} (\lambda  \varphi )} \sqrt{\alpha }\right){}^2\\
\notag & -\alpha  C_1^2 \lambda ^2 M_p^4 e^{-\lambda \varphi } I_{-2
\beta -1}\left(2 e^{-\frac{1}{2} (\lambda  \varphi )} \sqrt{\alpha
}\right) I_{1-2 \beta }\left(2 e^{-\frac{1}{2} (\lambda \varphi )}
\sqrt{\alpha }\right)+3 C_1^2 M_p^2 I_{-2 \beta }\left(2
e^{-\frac{1}{2} (\lambda  \varphi )} \sqrt{\alpha }\right){}^2\, .
\end{align}
The behavior of the Hubble rate as a function of the cosmic time can
only be found numerically, by finding the solution $\varphi (t)$, so
by solving numerically the differential equation (\ref{extra1}) for
$\varphi (0)=M_p$, in Fig. \ref{hubplot123} we can see that for a
quite large era the Hubble rate is almost constant, so it approaches
a de Sitter solution.
The solution (\ref{generalsolution1thrylos}) is a stable attractor
as our numerical analysis shows in Fig. \ref{hargarthrylos}, where
we used the values for the parameters as in Eq.
(\ref{choiceofparameterslocal}), and also the initial conditions,
$\varphi (0)=10M_p$ (red line), $\varphi (0)=M_p$ (green line),
$\varphi (0)=0.1 M_p$ (blue line).
What we are mainly interested for in this section is the behavior of
the linear perturbations $\theta (\varphi)$, the evolution of which
is described by the differential equation
(\ref{equationperturbations2}).
\begin{figure}[h]
\centering
\includegraphics[width=15pc]{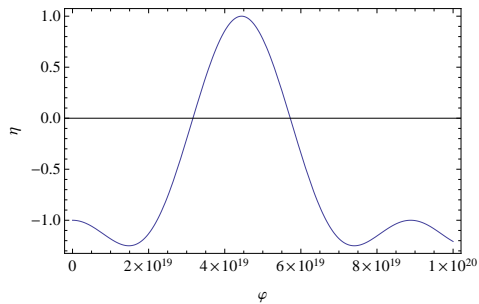}
\caption{$\varphi$-dependence of the function $f(\varphi)=-
\left(\sin ^2(\lambda  \varphi)+\cos (\lambda
\varphi)\right)$.}\label{asxetious}
\end{figure}
Without getting into many details, the first constant-roll era,
which corresponds to $\varphi/M_p\simeq 20$, is quite unstable as it
was expected, since the function $f(\varphi)$ for these field
values, is not slowly varying. However as the field value decreases,
the function $f(\varphi)$ approaches its limiting value
$f(\varphi)\sim -(\alpha+\beta^2)$, which is quite stable. Indeed
this can be verified for various initial conditions of $\theta
(\varphi)$, and the resulting evolution of perturbations is zero for
$\varphi/M_p\preceq 0.01$.
\begin{figure}[h]
\centering
\includegraphics[width=15pc]{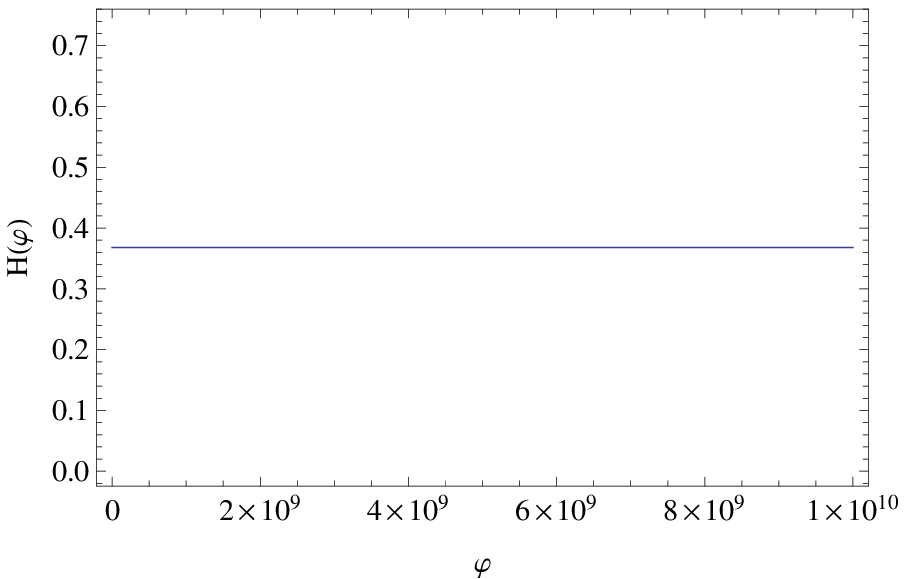}
\includegraphics[width=15pc]{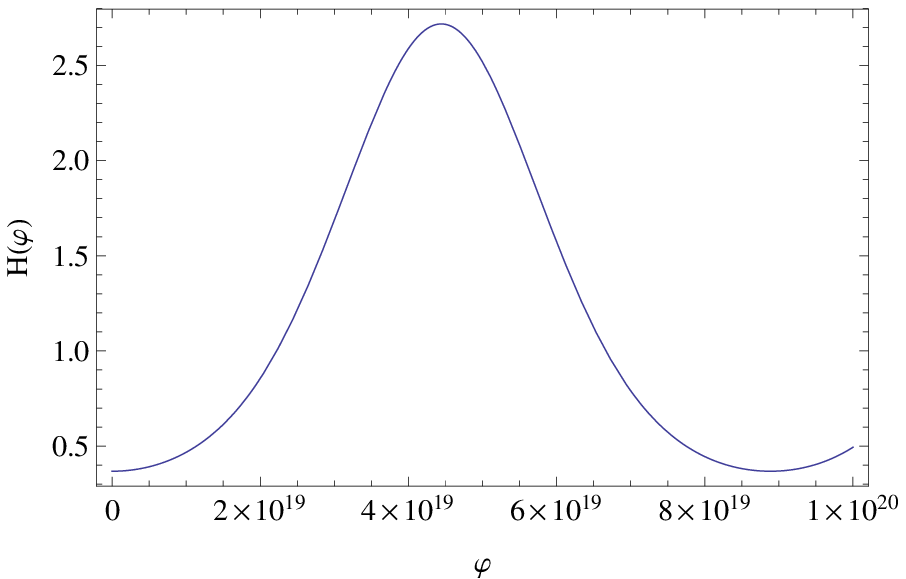}
\includegraphics[width=15pc]{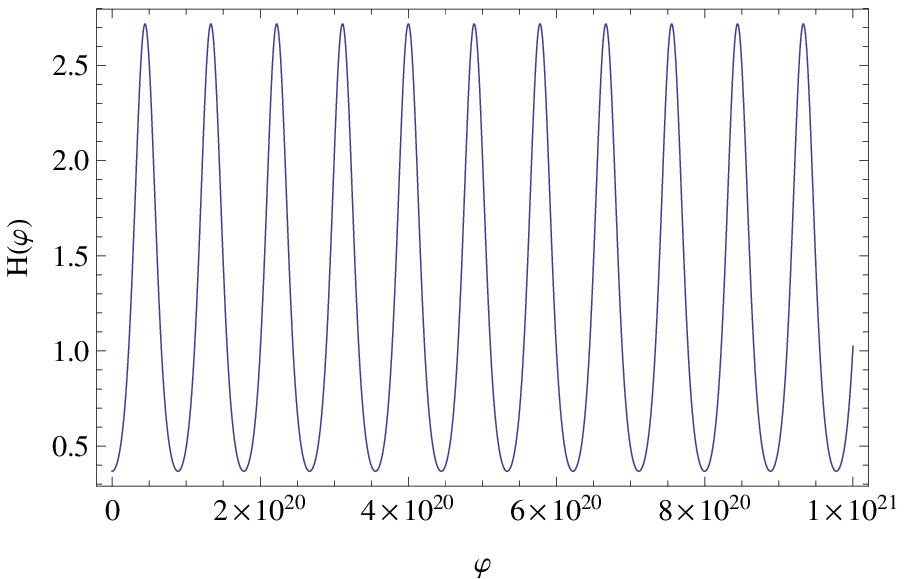}
\caption{$\varphi$-dependence of the Hubble rate $H(\varphi)=C_1e^{
(-\cos (\lambda  \varphi ))}$, for various scalar field
intervals.}\label{hargar}
\end{figure}
 In effect, this constant-roll era is also
haunted by the graceful exit issue, however, if the primordial
perturbations are generated during this era, which can be insured if
the preceding constant-roll era lasts only a few $e$-folds, then,
the resulting power spectrum is
\cite{Martin:2012pe,Motohashi:2014ppa},
\begin{equation}\label{pzthrylos}
\mathcal{P}_{\mathcal{R}}(k)\sim k^{2(-\alpha-\beta^2+3)}\, ,
\end{equation}
so the resulting spectral index is,
\begin{equation}\label{ns1thrylos}
n_s=1+2(-\alpha-\beta^2+3)\, .
\end{equation}
For the values of the parameters chosen as in Eq.
(\ref{choiceofparameterslocal}), the resulting spectral index is
$n_s\simeq 0.966$, so it is in agreement with the Planck data
\cite{planck}. Therefore this particular model shows that a viable
phenomenology may be provided by constant-roll to constant-roll
transitions. However, this model cannot be considered as being a
completely viable, since there are various important theoretical
issues that need to be amended, like the exit issue, the initial
state issue and the horizon crossing issue. The complete study of
these issues, exceeds the introductory aim of this work.
\begin{figure}[h]
\centering
\includegraphics[width=15pc]{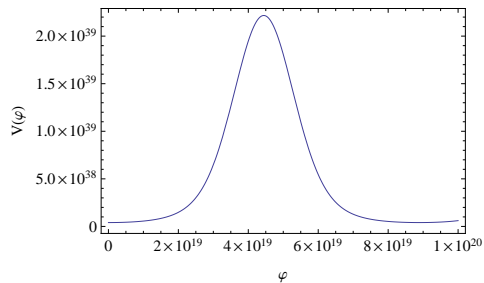}
\caption{$\varphi$-dependence of the scalar potential
(\ref{fullpotentialfordifficultcase}).}\label{potplot}
\end{figure}
In principle, the constant-roll to constant-roll transitions may
have an oscillatory behavior between constant-roll era. In the next
section we briefly discuss a toy model that exhibits this kind of
behavior.

\section{A Model Describing Oscillations Between Constant-Roll Eras}

Consider the following model for the function $f(\varphi)$,
\begin{equation}\label{choice2}
f(\varphi )=- \left(\sin ^2(\lambda  \varphi)+\cos (\lambda \varphi )\right)\, ,
\end{equation}
where $\lambda$ is given in Eq. (\ref{yperif}). It is obvious from
the form of the function (\ref{choice2}), that it can take bounded
values, due to the presence of the trigonometric functions, and also
it can be equal to zero. Hence, the cosmological system experiences
transitions between the slow-roll and constant-roll era, and also
from constant-roll to constant-roll. In order to have a concrete
idea of the behavior of $f(\varphi)$, in Fig. \ref{asxetious} we
plot the $\varphi$-dependence of $f(\varphi)$ for field values in
the interval $\varphi=(0,10 M_p)$. As it can be seen, the second
slow-roll index $\eta=-f(\varphi )$ oscillates between constant and
slow-roll eras, if of course the resulting Hubble rate is an
inflationary attractor. This is what we will investigate now.
\begin{figure}[h]
\centering
\includegraphics[width=15pc]{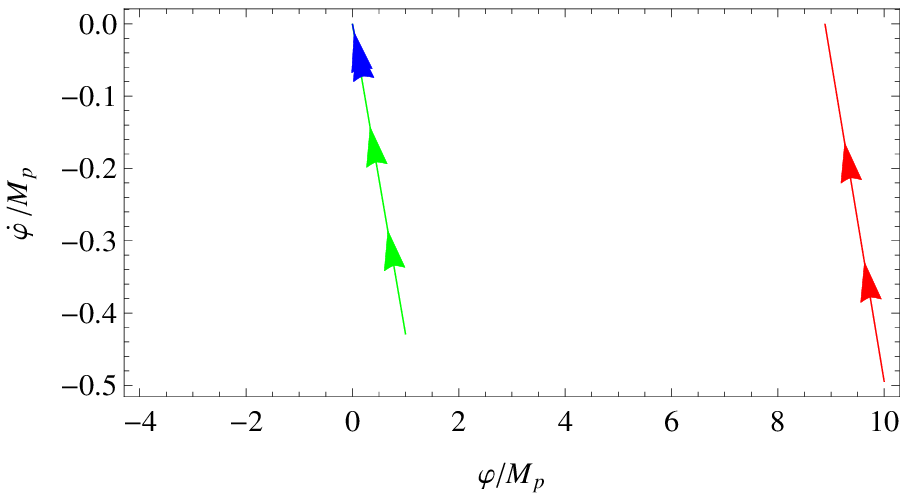}
\caption{Phase space structure of the solution
(\ref{hubbleratechoice2sol}), with the following initial conditions:
$\varphi (0)=10 M_p$ (red line), $\varphi (0)=M_p$ (green line),
$\varphi (0)=0.1M_p$ (blue line).}\label{akyroplot}
\end{figure}
The solution of the differential equation (\ref{masterequation}),
for the function $f(\varphi)$ chosen in Eq. (\ref{choice2}), is
given below,
\begin{equation}\label{hubbleratechoice2sol}
H(\varphi)=C_1e^{ (-\cos (\lambda  \varphi))}\, ,
\end{equation}
and in Fig. \ref{hargar} we plot the $\varphi$-dependence of the
Hubble rate $H(\varphi)$ given in Eq. (\ref{hubbleratechoice2sol}).
As it can be seen, the oscillatory behavior occurs in the Hubble
rate too, and the Hubble rate goes from a de Sitter vacuum to a de
Sitter vacuum, which correspond to the minima and maxima of the
function $\cos (\lambda  \varphi )$.
\begin{figure}[h]
\centering
\includegraphics[width=15pc]{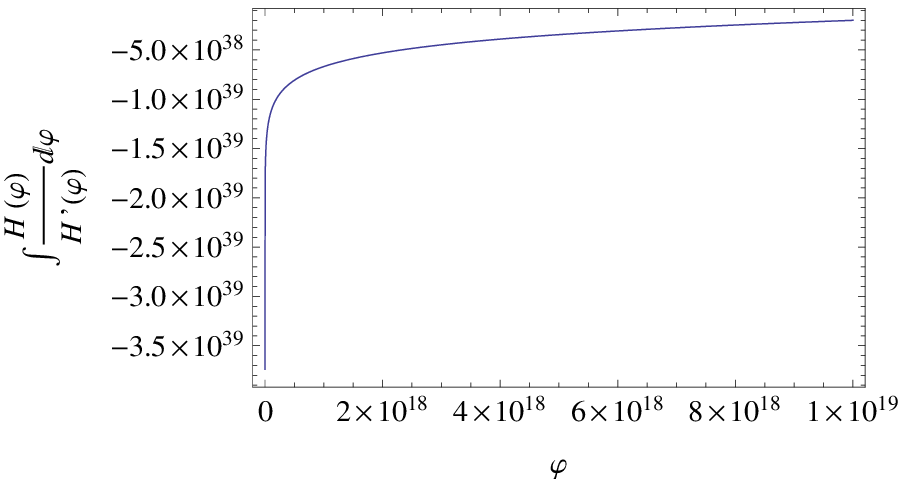}
\includegraphics[width=15pc]{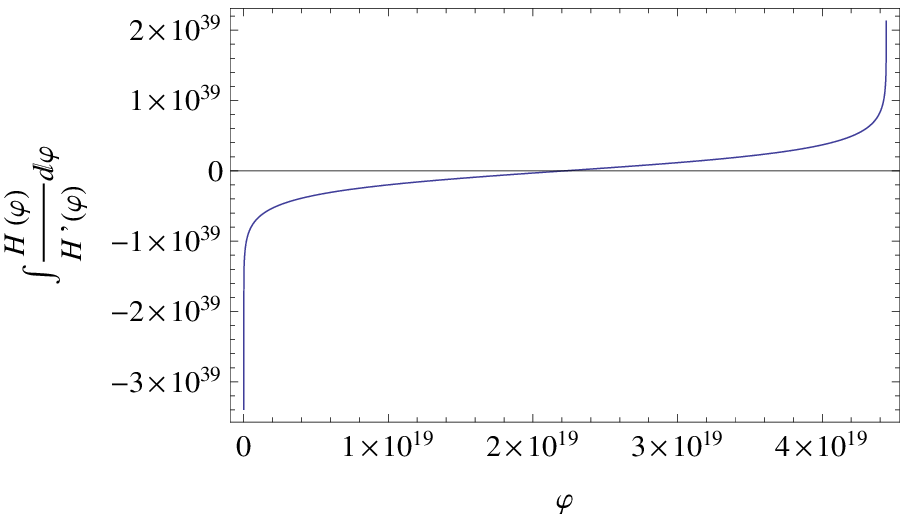}
\includegraphics[width=15pc]{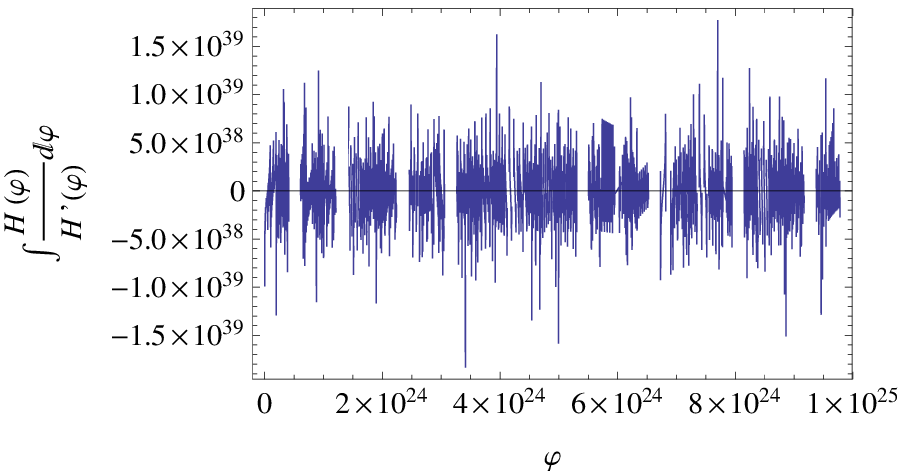}
\caption{$\varphi$-dependence of the linear perturbations $\theta
(\varphi)$ for the solution $H(\varphi)=C_1e^{ (-\cos (\lambda
\varphi))}$.}\label{ougk}
\end{figure}
By substituting Eq. (\ref{hubbleratechoice2sol}) in Eq.
(\ref{potentialhubblscalar}) we can find the potential which is,
\begin{align}\label{fullpotentialfordifficultcase}
& V(\varphi)=\frac{1}{2} C_1^2 M_p^2 e^{-2 \cos \left(\frac{\varphi
}{\sqrt{2} M_p}\right)} \left(\cos \left(\frac{\sqrt{2} \varphi
}{M_p}\right)+5\right)\, .
\end{align}
In Fig. \ref{potplot} we plot the potential
(\ref{fullpotentialfordifficultcase}) for $\varphi=(0,10M_p)$. We
need to investigate the stability properties of the solution
$H(\varphi )$ appearing in Eq. (\ref{hubbleratechoice2sol}), and in
Fig. \ref{akyroplot} we present the results of our numerical
analysis. We use three different initial conditions, namely,
$\varphi (0)=10M_p$ (red line), $\varphi (0)=M_p$ (green line),
$\varphi (0)=0.1M_p$ (blue line). As it can be seen, for the first
initial condition, the attractor is different in comparison to the
other two initial conditions. Oddly, after $\varphi \sim 10 M_p$,
the system develops peculiar properties, and becomes very unstable.
This can also be seen in Fig. \ref{ougk}, where we plot the behavior
of the linear perturbations (\ref{perturbationsolution1}), and as it
can be seen, after $\varphi \sim M_p$, the perturbations grow, and
after $\varphi \sim 10 M_p$, they develop a quite intense
oscillatory behavior. Therefore, the model (\ref{choice2}) after
$\varphi \sim M_p $ becomes unstable, however for smaller field
values, the solution (\ref{hubbleratechoice2sol}) seems to be the
attractor of the cosmological system. Hence, the model
(\ref{choice2}), has an odd behavior for large field values, however
for values $\varphi \preceq M_p$, it has interesting phenomenology,
since it allows oscillations between different roll eras.
Particularly it allows transitions between constant-roll eras and
slow-roll eras. A refined model of this sort could potentially have
an interesting phenomenology, but we defer this study to a future
work.

\section{Concluding Remarks}

In this paper we presented various models which allow transitions
between constant-roll eras. This was achieved by assuming that the
second slow-roll index $\eta$ is equal to a function of the scalar
field $f(\varphi )$. Then, by appropriately choosing the function,
it is possible to produce transitions between constant-roll eras,
and also it is possible to provide more involved transition
scenarios. We mainly focused on the stability behavior of the
solutions $H(\varphi)$ and we did not study in detail the
phenomenological implications of the models studied. As we
demonstrated, one of the models we studied, has very good stability
properties and also the solution $H(\varphi)$ is the attractor of
the cosmological equations. Another model we studied, which produces
oscillating transitions between constant-roll eras, is stable for a
$\varphi \preceq M_p$, so it has limited interest in comparison to
the other two models we presented.

In the class of exponential models we studied, we demonstrated that
the constant-roll eras are realized in the large and small field
values limits. As we demonstrated, the first constant-roll era is
unstable and the final constant-roll era is stable towards linear
perturbations of the constant-roll condition. This raises the
question with regards to the graceful exit, and if it is possible
for the system to finally exit from inflation. The lack of
analyticity in some cases makes it difficult to quantify the
slow-roll indices in terms of the $e$-foldings number, but in
principle the most interesting scenario is to combine the
constant-roll era(s) with a slow-roll era, in the spirit of Ref.
\cite{Odintsov:2017yud}, however the system should result to a
slow-roll era. In this way the slow-roll era could end in the usual
way it does in single scalar theories, when the slow-roll indices
become of order one. This feature is also an indication for the
graceful exit from inflation. Another issue is the calculation of
the power spectrum during the constant-roll eras, in which we used
standard approaches. However, an issue regarding the calculation is
the horizon crossing, since the power spectrum is not calculated at
the horizon crossing, so this is different in spirit in comparison
to the ordinary slow-roll case. As we showed by using the same
approach as in the related literature, the resulting power spectrum
can be nearly scale invariant. In conclusion, the virtue of having
two constant-roll eras is that the theory is more free from
unnecessary fine tunings, however the issue of the graceful exit
persists. We believe that the best proposal in this context is to
have a constant-roll era that ends up to a slow-roll era with the
mechanism we presented in this work. In this way, the theory
develops non-Gaussianities, which can be potentially measured by
future observations, but also the same theory can yield
observational indices corresponding to the slow-roll era.
Accordingly, the exit from inflation can come as an outcome of the
ending of the slow-roll era.

In a future work we shall address the phenomenological implications
of the models presented in this paper, mainly focusing on the
non-Gaussianities that can be produced during the constant-roll
eras. Of course it is conceivable that the models we presented
cannot be considered viable unless these produce results compatible
with the observational data, and also these models must solve
simultaneously the graceful exit issue and the non-Gaussianities
issue. However, our aim was not to provide viable models to all
aspects, but to present models that allow transitions between
constant-roll eras. Finding a viable model with these features is of
course the goal, but a more refined choice for the function
$f(\varphi)$ is required, and also the combination of an initial
constant-roll era that ends up to a slow-roll era seems the most
favorable scenario.

\section*{Acknowledgments}

This work is supported by MINECO (Spain), project
 FIS2013-44881, FIS2016-76363-P and by CSIC I-LINK1019 Project (S.D.O) and by Min. of Education and Science of Russia (S.D.O
and V.K.O).

\end{document}